\documentclass[12pt]{article}
 \usepackage{graphicx,amsfonts,amsmath}

 \renewcommand{\title}[1] {%
 \begingroup\begin{center}\vspace{0.0cm}\bf\Large
 \addtolength{\baselineskip}{1mm} #1 \end{center}\endgroup}

 \renewcommand{\author}[1] {%
 \begingroup\begin{center}\vspace{0.2cm}\bf #1 \vspace{0.2cm}
 \end{center}\endgroup}

 \newcommand{\address}[1] {%
 \begingroup\begin{center} #1 \end{center}\endgroup}

\begin{document}
 \title{Nonlinear differential equations
  \\ for the correlation functions   of the  2D Ising model \\ on the
 cylinder}
 \author{O. Lisovyy$^{\;*,\;\dag}$}
 \address{
  $^{*\;}$Bogolyubov Institute for Theoretical Physics \\
  Metrolohichna str., 14-b, Kyiv-143, 03143, Ukraine \vspace{0.2cm} \\
  $^{\dag\;}$ Laboratoire de Math\'ematiques et Physique Th\'eorique CNRS/UMR 6083,\\
  Universit\'e de Tours, Parc de Grandmont, 37200 Tours, France}
  \date{}

 \begin{abstract}
 We derive determinant representations and nonlinear differential
 equations for the scaled 2-point functions of the 2D Ising model
 on the cylinder. These equations generalize well-known results on
 the infinite lattice (Painlev\'e III equation and the equation for
 the $\tau$-function of Painlev\'e V).
 \end{abstract}

 \section{Introduction}

 One of the most beautiful results of the modern mathematical
 physics is the connection between correlations in the 2D Ising
 model and Painlev\'e functions. Quantum field theory and condensed
 matter physics usually deal with Dyson equations for the Green
 functions. These equations, however, can not be reduced to closed
 form, i. e. to find a propagator from Dyson equations one must
 first perturbatively calculate the vertex. The Ising model
 provides a remarkable example of closed differential equations for
 the correlation functions and, moreover, these equations prove to
 be integrable.


 Scaled 2-point function of the 2D Ising model on the infinite
 lattice
 $$ \langle\sigma(0,0)\sigma(r_{x},r_{y})\rangle^{(-)}=\xi\,\tau_{-}(t),
 \;\;\;s>1\eqno(1.1)$$
 $$ \langle\sigma(0,0)\sigma(r_{x},r_{y})\rangle^{(+)}=\xi\,\tau_{+}(t),
 \;\;\;s<1\eqno(1.2)$$
 can be expressed through the functions $\tau_{\pm}(t)$ of the
 single parameter
 $$ t=\frac{\left|s-1\right|}{\sqrt2}\sqrt{r_{x}^{2}+r_{y}^{2}},\eqno(1.3)$$
 where
 $$ s=\sinh 2\mathcal{K},\;\;\;\xi=\left|1-s^{-4}\right|^{\frac{1}{4}}\eqno(1.4)$$
 and $\mathcal{K}$ is the Ising coupling constant.

 As it was shown in \cite{mtw,pain3}, the function
 $$\eta(\theta)=\frac{\tau_{-}(2\theta)-\tau_{+}(2\theta)}{\tau_{-}(2\theta)+\tau_{+}(2\theta)}\eqno(1.5)$$
 satisfies the equation
 $$\eta''=\frac{1}{\eta}\left(\eta'\right)^{2}-\frac{1}{\theta}\,\eta'+\eta^{3}-\eta^{-1},\eqno(1.6)$$
 and the asymptotic boundary conditions
 $$\eta(\theta)\sim 1-\frac2\pi K_0(2\theta)\eqno(1.7)$$
 as $\theta\rightarrow\infty$. The equation (1.6) is a
 particular case of Painlev\'e III equation. Furthermore, if
 we define
 $$ \zeta(t)=t\frac{d\,\ln\tau_{\pm}(t)}{d t},\eqno(1.8)$$
 then in both cases \cite{pain5}
 $$\left(t\zeta''\right)^{2}=4\left(t\zeta'-\zeta\right)^{2}-4\left(\zeta'\right)^{2}\left(t\zeta'-\zeta\right)
 +\left(\zeta'\right)^{2}.\eqno(1.9)$$
 This is the equation for the $\tau$-function of Painlev\'e V. The
 most interesting feature of (1.9) is that it contains
 correlation functions $\tau_{\pm}$ themselves instead of their
 combination (1.5).

 There are several similar results for the correlations in
 one-dimensional quantum models: impenetrable Bose gas
 \cite{jimbo}, XY and XXZ spin chains \cite{perk, korepin},
 sine-Gordon field model \cite{leclair}. In two dimensions, there
 were no examples of this kind other than spin-spin correlation
 functions of the  Ising model on the infinite plane. Quite
 recently \cite{bugrij1, we} the form factor representation for
 the correlation function was calculated for the Ising model on a
 cylinder. This gave an opportunity to compute the susceptibility
 and to investigate its singularity structure in the complex
 temperature plane in the case of a finite-size lattice. In the
 present paper, we continue the study of correlations in the Ising
 model on a cylinder.

 In this case, the rotational invariance is broken even in the
 scaling limit and one has three scaling parameters:
 $$x=\frac{\left|1-s\right|}{\sqrt2}r_{x},\;\;\;y=\frac{\left|1-s\right|}{\sqrt2}
 r_{y},\;\;\;\beta=\frac{\left|1-s\right|}{\sqrt2}N,$$
 where $N$ is the number of sites on the base of the cylinder.
 Correlation function in the scaling limit is given by
 $$\langle\sigma(0,0)\sigma(r_{x},r_{y})\rangle^{(-)}=\xi\widetilde{\xi}_{T}(\beta)
 \mathrm{e}^{-|x|/\widetilde{\Lambda}(\beta)}\,\tau_{-}(x,y,\beta), \;\;\;s>1,$$
 $$\langle\sigma(0,0)\sigma(r_{x},r_{y})\rangle^{(+)}=\xi\widetilde{\xi}_{T}(\beta)
 \mathrm{e}^{-|x|/\widetilde{\Lambda}(\beta)}\,\tau_{+}(x,y,\beta), \;\;\;s<1.$$
 The notation will be explained in the next sections. Before
 summarize the main results of the paper, we define complex variables
 $$z=-\,\frac{|x|-i y}{2},\;\;\;\overline{z}=-\,\frac{|x|+i y}{2}\eqno(1.10)$$
 and the function
 $$\varphi=\ln\frac{\tau_{-}+\tau_{+}}{\tau_{-}-\tau_{+}}.\eqno(1.11)$$
 It will be shown that $\varphi$ satisfies the sinh-Gordon equation
 $$\varphi_{z\overline{z}}=\frac{1}{2}\sinh 2\varphi.\eqno(1.12)$$
 Besides that, we will derive the  equations for $\tau_{\pm}$
 separately. If we define $$u=\ln\tau_{\pm},\eqno(1.13)$$
 then $u$ satisfies the equation
 $$u_{z\overline{z}}\left(u_{zz\overline{z}\overline{z}}+2u_{z\overline{z}}^{2}-
 u_{z\overline{z}}\right)=
 u_{zz\overline{z}}u_{z\overline{z}\overline{z}}-u_{zz}u_{\overline{z}\overline{z}}\eqno(1.14)$$
 in both cases. This is quite a fascinating result. In fact, we
 will find a general $N$-solitonic solution of (1.14) and the
 Ising scaled correlation function is an infinite-solitonic one. As
 far as the author is aware, the equation (1.14) is not one of
 common knowledge. Nevertheless, it seems to be integrable and
 complete investigation of it (construction of the Lax
 representation, hamiltonian structure, etc.) is a challenging
 task.

 This paper is organized as follows. The scaling limit of the
 formulas of \cite{bugrij1,we} and determinant representations for
 the correlation functions on a cylinder are obtained in Section 2.
 On the infinite lattice correlation functions can be expressed
 through Fredholm determinants of integral operators and on the
 cylinder --- through the determinants of some infinite dimensional
 matrices. The derivation of the differential equations (1.12), (1.14)
 and a brief rederivation of Painlev\'e III (1.6) and Painlev\'e V
 (1.9) equations are given in Section~3.
 \section{Scaling limit and determinant representations\\ for the
 correlation functions}
 The most convenient representations for the investigation of
 correlation functions on the infinite lattice are the so-called
 form factor expansions \cite{palmer, yamada}:
 $$\langle\sigma(0,0)\sigma(r_{x},r_{y})\rangle^{(-)}=\xi\sum\limits_{n=0}^{\infty}
 g_{2n},\;\;\;\;\;s>1\eqno(2.1)$$
 $$\langle\sigma(0,0)\sigma(r_{x},r_{y})\rangle^{(+)}=\xi\sum\limits_{n=0}^{\infty}
 g_{2n+1},\;\;s<1\eqno(2.2)$$
 $$g_{n}=\frac{1}{n!(2\pi)^{n}}\int\limits_{-\pi}^{\pi}\ldots\int\limits_{-\pi}^{\pi}
 dq_{1}\ldots dq_{n}\,F_{n}^{2}[q]\prod\limits_{j=1}^{n}
 \frac{\mathrm{e}^{-|r_{x}|\gamma(q_{j})+ir_{y}q_{j}}}{\sinh \gamma(q_{j})},
 \;\;\;g_{0}=1,\eqno(2.3)$$
 $$F_{n}[q]=\prod\limits_{1\leq i<j\leq n}
 \frac{\sin\frac{q_{i}-q_{j}}{2}}{\sinh\frac{\gamma(q_{i})+\gamma(q_{j})}{2}},\;\;\;F_{1}=1,\eqno(2.4)$$
 where the function $\gamma(q)$ is defined by the equation
 $$\cosh\gamma(q)=s+s^{-1}-\cos q$$
 and condition $\gamma(q)>0$.

 It proved \cite{bugrij1, we} that the correlation function on the
 cylinder with $N$ sites on the base may be written in a similar
 form, namely
 $$\langle\sigma(0,0)\sigma(r_{x},r_{y})\rangle^{(-)}=\left(\xi\cdot\xi_{T}\right)\,
 \mathrm{e}^{-\left|r_{x}\right|/\Lambda}
 \sum\limits_{n=0}^{[N/2]}g_{2n},\;\;\;\;\;s>1\eqno(2.5)$$
 $$\langle\sigma(0,0)\sigma(r_{x},r_{y})\rangle^{(+)}=\left(\xi\cdot\xi_{T}\right)\,
 \mathrm{e}^{-\left|r_{x}\right|/\Lambda}
 \sum\limits_{n=0}^{[(N-1)/2]}g_{2n+1},\;\;\;\;s>1\eqno(2.6)$$
 $$g_{n}=\frac{\mathrm{e}^{-n/\Lambda}}{n!N^{n}}{\sum\limits_{[q]}}^{(b)}F_{n}^{2}[q]
 \prod\limits_{j=1}^{n}\frac{\mathrm{e}^{-\left|r_{x}\right|\gamma(q_{j})+ir_{y}q_{j}-\nu(q_{j})}}{\sinh\gamma(q_{j})},
 \;\;\;g_{0}=1. \eqno(2.7)$$

 The main differences between (2.1)--(2.3) and (2.5)--(2.7) are the
 substitution of the integration by the summation over discrete
 bosonic spectrum of quasimomentum ($q_{j}=\frac{2\pi}{N}\,k$,
 $k=0,\;1,\ldots,\;N-1$) and the appearance of cylindrical
 parameters $\xi_{T}$, $\Lambda$ and $\nu(q)$
 $$\ln\xi_{T}=\frac{N^{2}}{2\pi^{2}}\int\limits_{0}^{\pi}
 \frac{dp\; dq\, \gamma'(p)\gamma'(q)}{\sinh\left(N\gamma(p)\right)\sinh\left(N\gamma(q)\right)}
 \ln\left|\frac{\sin\left((p+q)/2\right)}{\sin\left((p-q)/2\right)}\right|,\eqno(2.8)$$
 $$\Lambda^{-1}=\frac{1}{\pi}\int\limits_{0}^{\pi}dp\,\ln\coth\left(N\gamma(p)/2\right),\eqno(2.9)$$
 $$\nu(q)=\frac{1}{\pi}\int\limits_{0}^{\pi}\frac{dp\,
 \left(\cos p-\mathrm{e}^{-\gamma(q)}\right)}{\cosh\gamma(q)-\cos p}
 \ln\coth\left(N\gamma(p)/2\right).\eqno(2.10)$$
 The finite number of terms in the sums (2.5) and (2.6) provides
 an opportunity of independent check of the formulas
 \cite{bugrij1, we} for small $N$.

 The scaling limit of the Ising model is of interest mainly due to
 its applications in quantum field theory. This limit implies that
 $$\gamma(0)\rightarrow 0,\;\;\;r_{x}\rightarrow\infty,\;\;\;r_{y}\rightarrow\infty,\;\;\;N\rightarrow\infty,$$
 $$\gamma(0)r_{x}\rightarrow x=\mathrm{const},\;\;\;\gamma(0)r_{y}\rightarrow
 y=\mathrm{const},\;\;\;\gamma(0)N\rightarrow \beta=\mathrm{const}.$$
 In this case \cite{bugrij2} the formulas (2.5)--(2.7) reduce to
 $$\langle\sigma(0,0)\sigma(r_{x},r_{y})\rangle^{(-)}=\xi\widetilde{\xi}_{T}(\beta)
 \mathrm{e}^{-|x|/\widetilde{\Lambda}(\beta)}\,\sum\limits_{n=0}^{\infty}\widetilde{g}_{2n},
 \;\;\;\;\;\;s>1,\eqno(2.11)$$
 $$\langle\sigma(0,0)\sigma(r_{x},r_{y})\rangle^{(+)}=\xi\widetilde{\xi}_{T}(\beta)
 \mathrm{e}^{-|x|/\widetilde{\Lambda}(\beta)}\,\sum\limits_{n=0}^{\infty}\widetilde{g}_{2n+1},
 \;\;\;s<1,\eqno(2.12)$$
 $$\widetilde{g}_{n}=\frac{1}{n!\beta^{n}}\sum\limits_{[l]=-\infty}^{\infty}\widetilde{F}_{n}^{2}[l]
 \prod\limits_{j=1}^{n}\frac{\mathrm{e}^{-|x|\sqrt{1+\left(\frac{2\pi l_{j}}{\beta}\right)^{2}}+
 iy\frac{2\pi l_{j}}{\beta}-\widetilde{\nu}(l_{j},\,\beta)}}
 {\sqrt{1+\left(\frac{2\pi l_{j}}{\beta}\right)^{2}}},\;\;\;\widetilde{g}_{0}=1,\eqno(2.13)$$
 $$\widetilde{F}_{n}[l]=\prod\limits_{1\leq i<j\leq n}\frac{\frac{2\pi l_{i}}{\beta}-\frac{2\pi l_{j}}{\beta}}
 {\sqrt{1+\left(\frac{2\pi l_{i}}{\beta}\right)^{2}}+\sqrt{1+\left(\frac{2\pi l_{j}}{\beta}\right)^{2}}}\;.\eqno(2.14)$$
 The summation over $[l]$
 $$\sum\limits_{[l]=-\infty}^{\infty}=\sum\limits_{l_{1}=-\infty}^{\infty}\ldots\sum\limits_{l_{n}=-\infty}^{\infty}$$
 is taken over integer values of $l_{j}$, and the quantities
 $\widetilde{\xi}_{T}(\beta)$, $\widetilde{\Lambda}(\beta)$,
 $\widetilde{\nu}(l,\beta)$ are defined by the formulas
 $$\ln\widetilde{\xi}_{T}\left(\beta\right)=\frac{\beta^{2}}{2\pi^{2}}\int\limits_{0}^{\infty}\int\limits_{0}^{\infty}
 \frac{p\;dp\;q\,dq\; \ln\left|\frac{p+q}{p-q}\right|}
 {\sqrt{p^{2}+1}\,\sqrt{q^{2}+1}\sinh\left(\beta\sqrt{p^{2}+1}\right)\sinh
 \left(\beta\sqrt{q^{2}+1}\right)},$$
 $$\widetilde{\Lambda}^{-1}(\beta)=\frac{1}{\pi}\int\limits_{0}^{\infty}dp\;\ln\coth\frac{\beta\sqrt{1+p^{2}}}{2},$$
 $$\widetilde{\nu}(l,\beta)=\frac{2}{\pi}\int\limits_{0}^{\infty}dp\;
 \frac{\sqrt{1+\left(\frac{2\pi l}{\beta}\right)^{2}}}{1+\left(\frac{2\pi l}{\beta}\right)^{2}+p^{2}}
 \ln\coth\frac{\beta\sqrt{1+p^{2}}}{2}.$$

 Let us denote
 $$u_{i}=\frac{2\pi l_{i}}{\beta}+\sqrt{1+\left(\frac{2\pi l_{i}}{\beta}\right)^{2}},\;\;\;
 {u_{i}}^{-1}=-\frac{2\pi l_{i}}{\beta}+\sqrt{1+\left(\frac{2\pi l_{i}}{\beta}\right)^{2}}.$$
 It is easily verified that
 $$\widetilde{F}_{n}^{2}[l]=\prod\limits_{1\leq i<j\leq n}\left(\frac{u_i-u_j}{u_i+u_j}\right)^{2}.$$
 The right hand side of the last formula can be expressed through
 the
 Vandermond determinant of a matrix with dimension $n\times n$:
 $$\mathrm{det}_n\left|\left|\frac{1}{u_i+u_j}\right|\right|=\prod\limits_{i=1}^{n}\left(\frac{1}{2u_i}\right)
 \prod\limits_{1\leq i<j\leq n}\left(\frac{u_i-u_j}{u_i+u_j}\right)^{2}. $$
 If we make use of this fact and denote
 $$E_i\equiv E_i(\beta)=\left[\frac{2u_i\mathrm{e}^{-\widetilde{\nu}(l_i,\,\beta)}}
 {\sqrt{1+\left(\frac{2\pi l_{i}}{\beta}\right)^{2}}}\right]^{\frac12},\eqno(2.15)$$
 then for $\widetilde{g}_n$ we obtain
 $$\widetilde{g}_n=\frac{1}{n!\beta^n}\sum\limits_{[l]=-\infty}^{\infty}
 \mathrm{det}_n\left|\left|\frac{E_i E_j
 \exp\left[-\frac{|x|-iy}{2}\,\frac{u_i+u_j}{2}-\frac{|x|+iy}{2}\,\frac{{u_i}^{-1}+{u_j}^{-1}}{2}\right]}
 {u_i+u_j}\right|\right|.\eqno(2.16)$$

 Let us consider an arbitrary square matrix $\widehat{K}$ with the
 elements $K_{mn}$ and construct the determinant
 $$\left|1+\frac{\widehat{K}}{\beta}\right|=\mathrm{e}^{\mathrm{Tr}\ln\left(1+\frac{\widehat{K}}{\beta}\right)}=
 \mathrm{e}^{\mathrm{Tr}\left(\frac{1}{\beta}\,\widehat{K}-\frac{1}{2\beta^2}\,\widehat{K}^2+\ldots\right)}=$$
 $$=1+\frac{1}{\beta}\,\mathrm{Tr}\widehat{K}+\frac{1}{2\beta^2}\,
 \left[\left(\mathrm{Tr}\widehat{K}\right)^2-\mathrm{Tr}\left(\widehat{K}^2\right)\right]+\ldots=$$
 $$=1+\frac{1}{\beta}\sum\limits_{m}K_{mm}+\frac{1}{2\beta^{2}}\sum\limits_{m,n}\left|
\begin{array}{cc}
  K_{mm} & K_{mn} \\
  K_{nm} & K_{nn} \
 \end{array}\right|+\ldots$$
 as the expansion over $\frac{1}{\beta}$. After comparing these
 expressions with (2.16), it is easy to see that
 $$\sum\limits_{n=0}^{\infty}\widetilde{g}_{n}=\left|1+\frac{1}{\beta}\,\widehat{K}\right|,$$
 $$\sum\limits_{n=0}^{\infty}(-1)^{n}{}\widetilde{g}_{n}=\left|1-\frac{1}{\beta}\,\widehat{K}\right|,$$
 where the elements of the infinite-dimensional matrix
 $\widehat{K}$ are obtained from
 $$K_{l_1 l_2}=\frac{E_1 E_2
 \exp\left[-\frac{|x|-iy}{2}\,\frac{u_1+u_2}{2}-\frac{|x|+iy}{2}\,\frac{{u_1}^{-1}+{u_2}^{-1}}{2}\right]}
 {u_1+u_2}\;,\eqno(2.17)$$
 and $l_1,\;l_2=-\infty,\ldots,0,\ldots,\infty$.

 Let us summarize: correlation functions of the Ising model on the
 cylinder in the scaling limit
 $$\langle\sigma(0,0)\sigma(r_{x},r_{y})\rangle^{(-)}=\xi\widetilde{\xi}_{T}(\beta)
 \mathrm{e}^{-|x|/\widetilde{\Lambda}(\beta)}\,\tau_{-}(x,y,\beta),
 \;\;\;T<T_{c},\eqno(2.18)$$
 $$\langle\sigma(0,0)\sigma(r_{x},r_{y})\rangle^{(+)}=\xi\widetilde{\xi}_{T}(\beta)
 \mathrm{e}^{-|x|/\widetilde{\Lambda}(\beta)}\,\tau_{+}(x,y,\beta),
 \;\;\;T>T_{c},\eqno(2.19)$$
 can be expressed through the determinants of infinite-dimensional
 matrices:
 $$\tau_{-}+\tau_{+}=\left|1+\frac{1}{\beta}\,\widehat{K}\right|,\eqno(2.20)$$
 $$\tau_{-}-\tau_{+}=\left|1-\frac{1}{\beta}\,\widehat{K}\right|,\eqno(2.21)$$
 where $\widehat{K}$ is defined by (2.17).

 On the infinite lattice $\widehat{K}$ reduces to integral
 operator with the kernel
 $$K(u,v)=\frac{1}{\pi}\,
 \frac{\mathrm{e}^{-\frac{|x|-iy}{2}\,\frac{u+v}{2}\,-\frac{|x|+iy}{2}\,\frac{u^{ -1}+v^{-1}}{2}}}{u+v},\eqno(2.22)$$
 acting on $L^2(0,\infty)$, and matrix determinants in the right
 hand side of (2.20) and (2.21) become Fredholm determinants.

 \section{Nonlinear differential equations}
 In recent years numerous connections between Fredholm
 determinants of certain integral operators and differential
 equations become apparent \cite{widom1, widom2, widom3}. We will
 formulate here only necessary results.

 Let us consider integral operator $\widehat{K}$ with the kernel
 $$K(x,y)=\frac{E(x)E(y)}{x+y}\,,\;\;\;\;E(x)=e(x)\,\mathrm{e}^{z\frac{x}{2}+\overline{z}\frac{x^{-1}}{2}}.$$
 It was shown in \cite{widom2} that the functions
 $$\varphi=\ln\left|1+\lambda\widehat{K}\right|-\ln\left|1-\lambda\widehat{K}\right|,\eqno(3.1)$$
 $$\psi=\ln\left|1+\lambda\widehat{K}\right|+\ln\left|1-\lambda\widehat{K}\right|,\eqno(3.2)$$
 satisfy the differential equations
 $$\partial_{z\overline{z}}\varphi=\frac12\sinh 2\varphi,\eqno(3.3)$$
 $$\partial_{z\overline{z}}\psi=\frac{1-\cosh 2\varphi}{2}.\eqno(3.4)$$
 To derive them, one has to define
 $$u_{i,j}=u_{j,\,i}=\langle E_i|\frac{1}{1-\widehat{K}^2}|E_j\rangle,\eqno(3.5)$$
 $$v_{i,j}=v_{j,\,i}=\langle E_i|\frac{\widehat{K}}{1-\widehat{K}^2}|E_j\rangle,\eqno(3.6)$$
 where $E_i(x)=x^i E(x)$, and to prove the following recursion
 relations
 $$u_{i+1,j}-u_{i,j+1}=u_{i,\,0}v_{j,\,0}-v_{i,\,0}u_{j,\,0},\eqno(3.7)$$
 $$v_{i+1,j}+v_{i,j+1}=u_{i,\,0}u_{j,\,0}-v_{i,\,0}v_{j,\,0}\eqno(3.8)$$
 and differentiation formulas
 $$ \Biggl\{\begin{array}{lll} \partial_z
 u_{p,\,q}=\frac12\,\left(u_{p,\,0}v_{q,\,0}+v_{p,\,0}u_{q,\,0}\right)+
 \frac12\left(u_{p+1,\,q}+u_{p,\,q+1}\right),  \\
 \partial_{\,\overline{z}}
 u_{p,\,q}=\frac12\,\left(u_{p,\,-1}v_{q,\,-1}+v_{p,\,-1}u_{q,\,-1}\right)+
 \frac12\left(u_{p-1,\,q}+u_{p,\,q-1}\right),  \end{array}\eqno(3.9) $$
 $$ \biggl\{\begin{array}{ll}
 \partial_z v_{p,\,q}=\frac12\,\left(u_{p,\,0}u_{q,\,0}+v_{p,\,0}v_{q,\,0}\right)+
 \frac12\left(v_{p+1,\,q}+v_{p,\,q+1}\right),  \\
 \partial_{\,\overline{z}} v_{p,\,q}=\frac12\,\left(u_{p,\,-1}u_{q,\,-1}+v_{p,\,-1}v_{q,\,-1}\right)+
 \frac12\left(v_{p-1,\,q}+v_{p,\,q-1}\right),  \end{array}\eqno(3.10) $$
 $$\partial_{z}\varphi=u_{0,0},\;\;\;\partial_{\,\overline{z}}\varphi=u_{-1,-1},\eqno(3.11)$$
 $$\partial_{z}\psi=-v_{0,0},\;\;\;\partial_{\,\overline{z}}\psi=-v_{-1,-1}.\eqno(3.12)$$
 Two additional equations
 $$\partial_{zz}\psi=-\left(\partial_{z}\varphi\right)^2,\eqno(3.13)$$
 $$\partial_{\,\overline{z}\overline{z}}\psi=-\left(\partial_{\,\overline{z}}\varphi\right)^2,\eqno(3.14)$$
 were not observed in \cite{widom2}. To derive, for example, first
 of them, one has to differentiate (3.12) with respect to $z$ and
 to use the recursion relation (3.8).

 It can be shown that the recursion relations (3.7), (3.8),
 differentiation formulas (3.9)--(3.12) and differential equations
 (3.3), (3.4), (3.13), (3.14) remain valid even when $\widehat{K}$
 is a matrix (finite- or infinite-dimensional) with the elements
 $$K_{mn}=\frac{e_m e_n \;\mathrm{e}^{\;z\frac{x_m+x_n}{2}+\overline{z}\frac{{x_m}^{-1}+{x_n}^{-1}}{2}}}
 {x_m+x_n}\;.$$
 The matrix (2.17) belongs to this class, so after denoting
 $$\varphi=\ln\frac{\tau_{-}+\tau_{+}}{\tau_{-}-\tau_{+}},\;\;\;\psi=
 \ln\left(\tau_{-}^{2}-\tau_{+}^{2}\right),\eqno(3.15)$$
 $$z=-\,\frac{|x|-i y}{2},\;\;\;\overline{z}=-\,\frac{|x|+i y}{2},\eqno(3.16)$$
 we can write down a system of equations for the correlation
 functions:
 $$\tau_{-}=\mathrm{e}^{\frac{\psi}{2}}\cosh\frac{\varphi}{2},\eqno(3.17)$$
 $$\tau_{+}=\mathrm{e}^{\frac{\psi}{2}}\sinh\frac{\varphi}{2},\eqno(3.18)$$
 $$\partial_{z\overline{z}}\varphi=\frac12\sinh 2\varphi,\eqno(3.19)$$
 $$\partial_{z\overline{z}}\psi=\frac{1-\cosh 2\varphi}{2},\eqno(3.20)$$
 $$\partial_{zz}\psi=-\left(\partial_{z}\varphi\right)^2,\eqno(3.21)$$
 $$\partial_{\,\overline{z}\overline{z}}\psi=-\left(\partial_{\,\overline{z}}\varphi\right)^2.\eqno(3.22)$$
 On the infinite lattice in the scaling limit correlation function
 and, therefore, the functions $\varphi$ and $\psi$ become
 invariant under rotations in the $(x,y)$ plane. In this case, one
 can define $\theta=2\left(z\overline{z}\right)^{\frac12}$,
 $\eta(\theta)=\mathrm{e}^{-\varphi(2z,\,2\overline{z})}$ and then sinh-Gordon equation (3.19) reduces to
 Painlev\'e~III equation (1.6) for the function $\eta(\theta)$.

 Before proceed with the differential equations, let us discuss the boundary conditions for (3.19). It
 should be noted that the asymptotic conditions at
 $|x|\rightarrow\infty$ in the case of the cylinder do not
 determine the solution uniquely. Indeed, the asymptotics of the functions
 $\tau_\pm(x,y)$ is determined by the first terms of the
 corresponding form factor expansions, and moreover, in the expression
 for $g_1$ (2.13) we should take $l=0$ (this is not the case for the
 infinite lattice, where the sum over $l$ transforms into an integral and
 gives the modified Bessel function). Therefore, we obtain
 $$\tau_+\rightarrow
 \frac1\beta\;\mathrm{e}^{-|x|-\widetilde{\nu}(0,\,\beta)}, \;\;\;\tau_-\rightarrow 1,$$
 as $|x|\rightarrow\infty$.
 The corresponding asymptotics of $\varphi$
 $$\varphi\rightarrow
 \frac2\beta\;\mathrm{e}^{-|x|-\widetilde{\nu}(0,\,\beta)}$$
 does not depend on $y$-coordinate. If we assume that the last
 condition defines the solution
 uniquely, it would not depend on $y$ for all values of
 $x$ and obviously, this is not true.

 If it were possible to obtain an additional differential equation,
 containing the derivatives with respect to $\frac1\beta$ (this
 quantity corresponds to QFT temperature), one would be able to use
 the solution of Painlev\'e III equation (1.6) with the boundary
 conditions (1.7) as the initial condition at $\frac1\beta=0$. However,
 the derivation of such equation remains an open problem. This problem is rather
 difficult to solve, mainly due to the complexity of the function
 $\widetilde{\nu}(l,\beta)$.

 Therefore, we have no way other than the classical one
 --- to determine, at least numerically, the function $\varphi(x,y)$
 and its derivative on the whole line ($y=\mathrm{const}$) or on the circle
 ($x=\mathrm{const}$). Such boundary conditions, combined with the
 periodicity in $y$-coordinate, will completely determine the solution
 of sinh-Gordon equation (3.19).

 Now we can turn to the derivation of the equations for the functions $\tau_{\pm}$. If one denotes
 $$u=\ln\tau_{-},$$
 then by means of usual differentiation and using the relations
 (3.17)--(3.22) it is straightforward to obtain the following
 identities:
 $$u_{zz}=\frac{\cosh\varphi-1}{2}\,\partial_{z}\left(\frac{\partial_{z}\varphi}{\sinh\varphi}\right),\eqno(3.23)$$
 $$u_{\overline{z}\overline{z}}= \frac{\cosh\varphi-1}{2}\,
 \partial_{\,\overline{z}}\left(\frac{\partial_{\,\overline{z}}\varphi}{\sinh\varphi}\right),\eqno(3.24)$$
 $$u_{z\overline{z}}=\frac{\cosh\varphi-1}{2}\,\left[-1+
 \left(\frac{\partial_{z}\varphi}{\sinh\varphi}\right)\left(\frac{\partial_{\,\overline{z}}\varphi}{\sinh\varphi}\right)
 \right],\eqno(3.25)$$
 $$u_{zz\overline{z}}=u_{z\overline{z}}\left(\frac{\partial_{z}\varphi}{\sinh\varphi}\right)+
 u_{zz}\left(\frac{\partial_{\,\overline{z}}\varphi}{\sinh\varphi}\right),\eqno(3.26)$$
 $$u_{z\overline{z}\overline{z}}=u_{\overline{z}\overline{z}}\left(\frac{\partial_{z}\varphi}{\sinh\varphi}\right)+
 u_{z\overline{z}}\left(\frac{\partial_{\,\overline{z}}\varphi}{\sinh\varphi}\right),\eqno(3.27)$$
 $$u_{zz\overline{z}\overline{z}}=u_{z\overline{z}\overline{z}}\left(\frac{\partial_{z}\varphi}{\sinh\varphi}\right)+
 u_{zz\overline{z}}\left(\frac{\partial_{\,\overline{z}}\varphi}{\sinh\varphi}\right)-
 2u_{z\overline{z}}^{2}+
 \frac{2\left[u_{zz}u_{\overline{z}\overline{z}}-u_{z\overline{z}}^2\right]}{\cosh\varphi-1}\;.\eqno(3.28)$$
 The last four relations can be treated as a system of algebraic
 equations for three variables $\frac{\cosh\varphi-1}{2}$,
 $\left(\frac{\partial_{z}\varphi}{\sinh\varphi}\right)$,
 $\left(\frac{\partial_{\,\overline{z}}\varphi}{\sinh\varphi}\right)$.
 The equation for correlation function
 $$u_{z\overline{z}}\left(u_{zz\overline{z}\overline{z}}+2u_{z\overline{z}}^{2}-u_{z\overline{z}}\right)=
 u_{zz\overline{z}}u_{z\overline{z}\overline{z}}-u_{zz}u_{\overline{z}\overline{z}}\eqno(3.29)$$
 appears as the consistency condition for this system. In the case
 $u=\ln\tau_{+}$ all the calculations can be performed in
 analogous fashion, and, though the equations (3.23)--(3.28) change
 their form, the equation (3.29) for the correlation function is
 the same. Since (3.29) is obtained from (integrable) sinh-Gordon
 equation (3.19) and three linear equations (3.20)--(3.22), it is
 very likely integrable. More detailed investigation of (3.29) is
 beyond the scope of this paper.

 On the infinite lattice the problem of derivation of differential
 equations for the correlation functions is much simpler due to
 rotational invariance. The functions $\tau_{\pm}$, $\varphi$,
 $\psi$, $u$ depend on the single variable
 $t=2\left(z\overline{z}\right)^{\frac12}$. Instead of (3.19) one has
 $$\varphi''+\frac{1}{t}\,\varphi'=\sinh\varphi\;\cosh\varphi,$$
 and instead of (3.23), (3.25), (3.26)
 $$u''-\frac{1}{t}\,u'=\frac{\left(\varphi''-\frac{1}{t}\,\varphi'\right)\sinh\varphi-\left(\varphi'\right)^2 \cosh\varphi}
 {2(\cosh\varphi+1)},$$
 $$u''+\frac{1}{t}\,u'=\frac{\cosh\varphi-1}{2}\,\left[-1+\left(\frac{\varphi'}{\sinh\varphi}\right)^2\right],$$
 $$\left(u''+\frac{1}{t}\,u'\right)'=2u''\left(\frac{\varphi'}{\sinh\varphi}\right),$$
 correspondingly. When we again regard the last four relations as
 a system for three variables $\varphi$, $\varphi'$, $\varphi''$
 and if we denote $$\zeta=tu',$$ we will obtain well-known result
 $$\left(t\zeta''\right)^{2}=4\left(t\zeta'-\zeta\right)^{2}-4\left(\zeta'\right)^{2}\left(t\zeta'-\zeta\right)
 +\left(\zeta'\right)^{2}$$
 --- the equation for the $\tau$-function of the Painlev\'e V
 equation, --- as the consistency condition for this system.
 Analogous calculations lead to the same equation for $\tau_{+}$.

 \vspace{1cm}

 I am grateful to V. N. Shadura for suggesting the problem and
 constant help, to V.~N.~Rubtsov and especially to A.~I.~Bugrij --- for fruitful discussions and
 critical comments. This work was supported by the INTAS program
 under grant INTAS-00-00055 and by Ukrainian SFFR project
 No. 02.07/00152.

 \bibliographystyle{plain}
 
 \end{document}